# A detailed analysis of impact collision ion scattering spectroscopy of bismuth selenide

Weimin Zhou, Haoshan Zhu, and Jory A. Yarmoff*

Department of Physics and Astronomy, University of California, Riverside, Riverside, CA 92521

**Abstract**

Impact collision ion scattering spectroscopy (ICISS), which is a variation of low energy ion scattering (LEIS) that employs large scattering angles, is performed on $Bi_2Se_3$ surfaces prepared by ion bombardment and annealing (IBA). ICISS angular scans are collected experimentally and simulated numerically along the $[120]$ and $[\bar{1}\bar{2}0]$ azimuths, and the match of the positions of the flux peaks shows that the top three atomic layers are bulk-terminated. A newly observed feature is identified as a minimum in the multiple scattering background when the ion beam incidence is along a low index direction. Calculated scans as a function of scattering angle are employed to identify the behavior of flux peaks to show whether they originate from shadowing, blocking or both. This new method for analysis of large-angle LEIS data is shown to be useful for accurately investigating complex surface structures.

---

*Corresponding author, E-mail: yarmoff@ucr.edu

# I. Introduction

Impact collision ion scattering spectroscopy (ICISS) is a variation of low energy ion scattering (LEIS) that is a simple and powerful way to investigate the atomic structure of the top few layers of a single crystal surface [1-3]. ICISS measures the intensity of projectiles singly scattered at a large fixed angle as the angle between the incident beam and the sample surface is scanned. ICISS is easy to perform and reveals information on the relative positions of atoms in real space, unlike diffraction based techniques that involve reciprocal space and require large amounts of computational time to solve a particular structure [4,5]. Flux peaks are observed in ICISS scans that result from projectiles that make a grazing collision with one atom and backscatter from a neighboring atom. The position of these flux peaks provides structural information for the top 2 to 4 atomic layers. Note that the popular technique, scanning tunneling microscopy (STM), provides information on the only outermost surface layer [6].

LEIS uses ion projectiles with kinetic energies on the order of keV [7,8]. In this range, the projectile energy is much larger than atomic bonding energies and the scattering cross sections are smaller than the interatomic distances, so that the interactions between the atoms in the solid can be ignored. Thus, the binary collision approximation (BCA) is valid, which means that LEIS trajectories can be treated classically as a sequence of individual projectile-target atom collisions. The scattered energy for a collision at a specific scattering angle depends primarily on the mass of the target atom, so that each elemental species on the surface produces a single scattering peak (SSP) in an energy spectrum. In this way, LEIS spectra directly provide the surface composition. This also means that the SSP energy for each element in the target is nearly constant during the collection of an ICISS scan since the scattering angle is fixed.

Previous [9-12] and upcoming publications from our group use LEIS and ICISS angular scans to investigate $Bi_2Se_3$ surfaces. $Bi_2Se_3$ belongs to the topological insulator (TI) class of materials, which is an emerging new state of quantum matter due to its unique topological surface states (TSS) [13-15]. $Bi_2Se_3$ is attractive for future applications because the TSS form a simple Dirac cone, its band gap has a practical value of 0.3 eV, and it is easy to grow single crystals. Single crystal $Bi_2Se_3$ consists of stacked quintuple layers (QL) ordered as Se-Bi-Se-Bi-Se [16]. The interaction between QLs is a weak van der Waals force, so that the samples naturally cleave between QLs leaving a Se-terminated surface, which is also referred to as a QL-termination or bulk-termination. The TSS exist completely within the uppermost QL [14,17], although there can be a contribution from the second uppermost QL for some surface structures [18-20]. Thus, a determination of the atomic structure of the outermost layers of $Bi_2Se_3$ is crucial to understand and control its novel electronic properties.

Our prior ICISS work [9,10] used polar scans along the [120] azimuth to show that $Bi_2Se_3$ surfaces prepared by ion bombardment and annealing (IBA) and *in situ* cleaving have similar atomic structures. The conclusion was reached by simply comparing the positions of the features in the ICISS polar scans, however, without explicitly identifying the source of each feature.

In the present study, ICISS polar scans along the [120] and [$\bar{1}\bar{2}0$] azimuths are collected from IBA-prepared surfaces and analyzed in detail to identify the trajectories responsible for each of the features and how they relate to the surface structure. In addition, a new method employing simulations of ICISS polar scans over a large range of scattering angles is introduced. This method enables the unambiguous identification of whether a feature is caused by shadowing, blocking or both. This approach is generally useful for analyzing experimental large angle LEIS angular scans for a variety of systems with a higher precision than can be achieved by simulations only along the scattering angle(s) used for data collection.

## II. Experimental Procedure

Single crystal $Bi_2Se_{3.12}$ was grown by melting a stoichiometric mixture of Bi and Se shot (99.999%, Alfa Aesar) in an evacuated 17 mm inner diameter quartz ampule following the slow cooling recipe described in Ref. [9]. The mixture was heated at 750ºC for one day, cooled to 500ºC at a rate of 3.7ºC $hr^{-1}$, and then annealed at 500ºC for three days. The ingots, which cleave naturally along the (001) plane, are broken into pieces around 10 mm in diameter. The samples are mounted onto transferable Ta sample holders by spot-welded Ta strips, cleaved a few times in air to obtain visually flat and shiny surfaces and then inserted into a load lock attached to an ultra-high vacuum (UHV) chamber.

Sample surface preparation and measurements are all conducted in this UHV chamber, which has a base pressure of $2\times10^{-10}$ Torr. The surfaces are prepared by 0.5 keV $Ar^+$ ion bombardment and annealing (IBA) as explicitly described elsewhere [10]. Briefly, samples are degassed at 130ºC for 2 hours, bombarded by 0.5 keV $Ar^+$ for 2 hours and annealed at 130ºC for 30 min to clean the surface. Samples are then recrystallized by repeated cycles of 30 min ion bombardment and 30 min annealing at 510ºC until a sharp and bright low energy electron diffraction (LEED) pattern is obtained.

LEIS is performed using a 3 keV $Na^+$ ion gun and a 160° Comstock AC-901 hemispherical electrostatic analyzer (ESA). The ESA, which is mounted in a fixed position, has a radius of 47.6 mm and 2 mm diameter entrance and exit apertures making the acceptance angle approximately 2°. The $Na^+$ ion gun is mounted on a turntable that can rotate around the chamber axis to adjust the scattering angle. The Na$^+$ beam current is typically 1 nA in a spot size approximately 1 mm in diameter. The foot of the sample manipulator allows for two rotational degrees of freedom. It can rotate about its own surface normal to change the azimuthal angle and it can also rotate about the chamber axis using a stepper motor to adjust the polar angle.

For the ICISS experiments reported here, the ion gun is kept stationary at a scattering angle of 161º with respect to the ESA, which is the maximum possible due to the size of the ion gun and analyzer. Before angular scans are performed, the ESA is used to collect an energy spectrum to determine the SSP energies for Se and Bi. To collect a LEIS angular scan, the detection energy of the ESA is fixed to continuously record the intensity of a particular SSP while the stepper motor automatically rotates the incident polar angle from 0º to 90º. Each ICISS polar scan takes about 100 seconds to collect, and different spots on the samples are used for each scan, so that the Na$^+$ ion beam fluence is kept below 1% of a monolayer at each spot to limit any damage to the sample surfaces. The ESA can only detect ions, but the percentage of Na$^+$ neutralized after scattering from Bi$_2$Se$_3$ is only around 5% [12], so that any change to neutralization with angle will not affect the features in the LEIS angular scans.

Molecular dynamics (MD) simulations of LEIS angular scans employing the BCA are performed using Kalypso [21]. The Thomas-Fermi-Molière repulsive potential using the Firsov screening length, corrected by a factor of 0.8 as determined in Ref. [11], is used in calculating each projectile-target atom interaction. The cut-off distance for this potential is 2.9 Å. The target model is a two-dimensional (2D) ($\bar{1}$20) plane that has three atomic layers ordered as Se-Bi-Se, unless stated otherwise. Periodic boundaries are applied parallel to the surface. The atomic positions in the target are taken as the average of the two sets of structural parameters determined by SXRD and LEED in Ref. [22]. The spacing between the top two atomic layers is set to 1.55 Å based on the work presented in Ref. [11]. The mean square vibrational amplitudes of the atoms in the bulk material are calculated using a Debye temperature of 200 K [23]. The vibrational amplitudes of atoms in the top two layers are isotropically enhanced by a factor of 4 by setting their Debye temperatures to 100 K. The acceptance angle used in the simulations is 2º to match that of the experiment.

**III. Results and Discussion**

Bi$_2$Se$_3$ surfaces prepared by IBA are bulk terminated and show a bright and sharp hexagonal LEED pattern, indicating a well-ordered surface [10]. The LEED pattern has a six-fold rotational symmetry, while the sample surface has a three-fold symmetry, however, so that the LEED patterns themselves are not sufficient to distinguish the [120] from [$\bar{1}\bar{2}0$] azimuth. Instead, LEIS is used to identify the specific orientation, as discussed below.

ICISS relies on projectiles undergoing a grazing collision with one atom, and then making a hard collision from a second atom [1-3]. Very little energy is lost in the grazing collision, while the hard collision leads to a scattered projectile at the SSP energy associated with the second atom. ICISS angular scans are then collected by monitoring the intensity of a particular SSP as the incident ion beam polar angle is adjusted relative to the surface plane. Flux peaks are observed in the ICISS scans that can be assigned to pairs of neighboring atoms in the crystal structure. The positions of the flux peaks are analyzed to reveal the full atomic structure of the outermost few layers.

To illustrate this process, Fig. 1 shows a string of identical atoms with arrows representing possible incoming and outgoing projectile trajectories with α being the polar angle of the incoming ions with respect to the atomic chain. Note that the arrows indicate the experimental scattering angle of 161°, which is formally defined as the change in angle of the trajectory caused by the collision. The curved dashed lines in this figure illustrate shadow cones, which are the regions behind each atom that the incoming projectiles cannot reach [8]. Shadow cones are formed by mapping out the trajectories expected from a parallel beam of incoming ions so that that the ends of each cone are parallel. The radii of shadow cones for low energy ions are on the order of Å [24]. For those trajectories that pass the atom after making a grazing collision, the ion flux is enhanced at the edges of the cones. When the ion beam incidence angle is nearly parallel to the chain, i.e., α is close to zero, all of the atoms are shadowed by the cones of their neighbors, so that no projectiles are able to make a hard collision and no backscattering occurs. As the sample is rotated with respect to the ion beam and α increases to a particular angle, called the critical angle, the edges of the cones all intersect their neighboring atoms. The orientation of the shadow cones and trajectories shown in Fig. 1 corresponds to such a critical angle. This results in a strong flux of projectiles impacting the neighboring atoms so that the backscattered yield is very large. As α is further increased so that the edges of the cones no longer interact with the neighboring atoms, the scattered ion yield becomes equivalent to the cross section for scattering from a row of isolated target atoms. In this way, an ICISS polar scan from a chain of atoms starts at a zero intensity, then builds to a flux peak at the critical angle, and finally settles down to a constant value when α is well past the critical angle.

Since a real single crystal consists of many chains of atoms at different angles with respect to the sample surface, flux peaks occur whenever the incident ion beam passes through an atomic chain. Since the flux peaks occur because of the interaction of two neighboring atoms in a chain, peaks in an ICISS scans can generally be modeled by sets of two atoms representing the outermost atoms in each chain. The critical angle for the chain depends on the shape of the shadow cone, which is dependent on the incident kinetic energy and the masses of the projectile and the first atom, as well as the orientation and distance between the two atoms. When a flux peak is due purely to the interaction between atoms in the outermost layer, it is called a surface flux peak (SFP). The SFP is the first feature that appears in an ICISS polar scan. Analysis of the positions of the flux peaks in an ICISS scan enables a determination of the atomic positions for a simple structure to be made that only requires a knowledge of the sizes and shapes of the shadow cones [3].

It is most common for ICISS experiments to utilize a scattering angle larger than 160º but smaller than 180º, however, which means that effects due to blocking of projectiles along the exit trajectory cannot be completely avoided. Note that ion scattering at exactly 180º is possible, but is not easy to achieve because of the need for a special MCP with a hole in the center [25]. Blocking cones are a similar concept as shadow cones, but are formed by ions exiting the surface after scattering from an atom in a deeper layer. Blocking cones have a similar size, but a different shape than shadow cones. The ends of a blocking cone form an angle as they arise from trajectories that radiate from a single atom. The flux at the edge of a blocking cone is enhanced in the same way as it is at the edge of a shadow cone.

The inset of the upper panel in Fig. 2 shows a representative energy spectrum collected at a 12º incident polar angle. The spectrum has clear Se and Bi SSPs that are well separated from each other. There is a step-like background tail of multiply scattered trajectories associated with each of the SSPs that continues at a constant value towards lower energies. The background associated with the Se SSP is about an order of magnitude larger than the background underneath the Bi SSP. This background of multiply scattered projectiles contributes intensity to each experimental SSP that is not pure single scattering, and this will occur to a greater extent for the Se SSP. Since the experimental ICISS scans measure the total intensity at the SSP energy, this background cannot be removed from the data. The background in the simulations, on the other hand, is largely absent, because the target model is composed of only a few atomic layers thereby strongly reducing the contribution of multiple scattering.

Figure 2 shows experimental and simulated ICISS polar scans collected along the [120] azimuth. There are six features in the experimental data in Fig. 2 that are labeled by a number and a symbol. The number is arbitrarily set to the order that the feature appears in the polar angle scans. The features are generally regarded as flux peaks with a maximum at a particular angle. It will be shown below that this is true for all of them except feature 3. The symbols "s" and "b" stand for "shadow" and "blocking", respectively, which indicates the type of cone from which the features are formed. The nature of the features is identified through the detailed analysis below. Note that features can result from a combination of shadowing and blocking, such as 2sb, while some can be caused by other phenomena.

Figure 3 displays six possible trajectories that could contribute to the features in Fig. 2, assuming that the sample is bulk-terminated. The cones pointing down represent shadow cones while the cones pointing up are blocking cones. The correspondence between the features and the trajectories are based on their angles and the same labels are used in both figures.

It is straightforward to assign features 1s, 4s, 5s, 6b as flux peaks because the atom pairs at the surface of the relevant atom chains are simple and the trajectories are clear. Feature "1s" in Fig. 2, for example, which is represented by the 1s trajectory in Fig. 3, occurs when the shadow cone of a first layer Se atom intersects the adjacent first layer Se atom, which forms the SFP in scattering from Se. Feature 4s is a peak observed in the Bi SSP ICISS scan that is caused by interaction of a first layer Se shadow cone with a second layer Bi atom. Similarly, feature 5s is a flux peak seen in the Se SSP scan that is caused by a second layer Bi shadow cone edge hitting a third layer Se atom. From the geometry of the crystal, as indicated in Fig. 3, it is expected that the polar angles for 4s and 5s should be close to each other. Because Bi is more massive than Se, however, the shadow cone caused by Bi is larger than that caused by Se so that 5s appears at a slightly larger polar angle than 4s, as verified by the experimental data in Fig. 2. Feature 6b in the Se SSP yield does not involve shadowing, but is caused as a projectile that backscatters from a third layer Se atom passes near of the edge of the blocking cone created by a first layer Se atom. Features purely due to blocking are commonly found at larger polar angles in ICISS scans collected with a scattering angle less than 180°.

The trajectories that lead to features 2sb and 3 in Fig. 2 are, however, more complicated than a simple two-atom analysis can explain. Their intensity is smaller than the other features, so it is possible that they involve both shadowing and blocking. A visual analysis of the crystal structure combined with simulations is used to assign the trajectories responsible for these features.

Based on the polar angle of feature 2sb, it is suspected to correspond to trajectory 2sb in Fig. 3, which involves three atoms in the crystal structure. In trajectory 2sb, the edge of the shadow cone of a first layer Se atom impacts a second layer Bi atom. The projectiles that hard scatter from the second layer Bi atoms then interact along the outgoing trajectory with the edge of the blocking cone created by the first layer Se atom that is adjacent to the original first layer Se atom. Thus, the alignment of these three atoms leads a small peak in the Bi SSP ICISS scan. A qualitative visual analysis cannot by itself substantiate whether or not trajectory 2sb actually leads to a flux peak nor whether the peak position matches that of feature 2sb. The simulations do, however, reveal a peak at the 2sb position with the correct intensity relative to feature 4s, thereby confirming the assignment.

The nature of feature 3 is even more complicated than feature 2sb, as it is not formed by an enhancement of the single scattering yield due to shadowing or blocking. At first, it might be suspected that feature 3 is similar in nature to feature 2sb expect that the hard collision involves a Se atom. If this were the case, then it would correspond to the three-atom trajectory 3sb shown in Fig. 3, which is caused by first-to-third layer shadowing followed by second-to-third layer blocking. The simulations shown in Fig. 2, however, do not reveal a peak between 20º and 40º, implying that trajectory 3sb does not create a flux peak. Note that simulations using a full QL still do not reveal a peak corresponding to feature 3. On the other hand, if the interlayer spacing between the second and third layers is increased from 1.95 to 2.05 Å to increase the probability for such a trajectory, the simulations do reveal a peak, but it is at 23º which is far off from the position of feature 3 in the experimental data. Thus, it appears that trajectory 3sb is not responsible for feature 3.

Instead, it is believed that feature 3 is formed in a novel manner by a change in the multiple scattering background intensity that the simulations do not reproduce. Inspection of Fig. 2 reveals that for all polar angles above about 15°, the experimental Se SSP intensity is much larger than the yield in the simulations, while the relative intensities of the features and background for the Bi SSP match well. It can thus be concluded that the multiple scattering background underneath the Se SSP leads to this extra yield in the experimental ICISS scan. Most of the background is fairly constant, so that the features 5s and 6s are clearly seen in both experiment and simulation. There is, however, no indication of a peak in the simulation that corresponds to feature 3, which provides additional evidence that it is not a flux peak.

Rather than considering feature 3 as a flux peak at 32º, it can instead be thought of as a one side of a minimum that occurs at approximately 37º. This angle corresponds to the [5 10 1] crystal direction along which there are channels between the atomic chains [16]. The distance between atom chains is larger than effective radii of the shadow cones. Thus, when the incident beam is aimed along this direction, ions can travel much deeper inside the crystal before making any collisions, and are therefore less likely to undergo multiple collisions that allow them to escape the surface with an energy close to that of the Se SSP. When the incident beam is a few degrees off of the [5 10 1] direction, then the incoming ions can collide with deeper lying atoms and initiate a multiple scattering trajectory that will lead to an emitted projectile. This effect is similar to, but different than "channeling" in high energy ion scattering [26], as channeling generally measures the yield of singly scattered ions. In the present case, the multiple scattering background decreases sharply at this particular incident angle leading to a minimum in the experimental scattered yield near 37°. The experimental Bi SSP ICISS scan also shows a small minimum near 37º, which could be due to a similar decrease of the multiple scattering background intensity. The minimum for the Bi SSP is not as obvious as for Se SSP because it has a much smaller background.

Figure 4 shows experimental and simulated ICISS scans of the Bi SSP collected along the [$\bar{1}\bar{2}0$] azimuth, which is 180° rotated from the [120] azimuth. The flux peak at 33º in both the experimental and simulated data is caused by the edge of a first layer Se atom shadow cone impacting a second layer Bi atom. The positions of the flux peaks are significantly different from those collected along the [120] azimuth. The correspondence of the experimental and calculated flux peak positions indicates that the choice of azimuths to represent the [120] and [$\bar{1}\bar{2}0$] directions, which cannot be ascertained solely from the LEED pattern, is correct.

In addition, the experimental data in Fig. 4 show a slow increase of Bi SSP intensity from 5º to 24º. Simulations using three layers fail to reveal any intensity in this region, but simulations using a full QL do show a peak between 13º and 22º, as shown in the figure. Thus, the intensity in this region has a contribution from trajectories that involve fourth layer Bi, although the specific trajectories are not identified. It can also be further inferred that the inclusion of even more layers in the simulations would show additional intensity at these low angles, presumably providing a better match to the experimental data. This illustrates a limitation of ICISS in that complex trajectories can contribute flux peaks to experimental data that a simple analysis does not reveal.

The positions of most the features in the experimental data and simulations in Figs. 2 and 4 are in general agreement, which verifies that the actual surface structure matches that of the model used in the calculations. The one peak that is not reproduced in the simulations, feature 3, is explained instead as one side of a minimum caused by a change in the multiple scattering background yield. This explanation is also consistent with the QL-terminated surface structure. Note that, in general, the Bi SSP simulations match the experimental data better because of the extra multiple scattering associated with the Se SSP.

Although such comparisons of experiment and simulation in an ICISS polar scan are useful for structural analysis, they do not directly show whether or not any particular feature is due to shadowing or blocking, or both. However, if data is also collected as a function of scattering angle, the following three principles can be used to clarify the nature of each flux peak. (1) The incident polar angle of features due purely to shadowing do not change with scattering angle because shadow cones are only determined by incident beam direction. (2) The incident polar angle of features due purely to blocking changes linearly with scattering angle because blocking cones only depend on the outgoing beam direction which does change with scattering angle. (3) The polar angle of features due to both shadowing and blocking do not change with scattering angle but their intensity changes sharply and goes to zero at a scattering angle corresponding to the bond direction between the final two atoms in the trajectory. This is because the enhanced intensity is primarily due to focusing at the edge of a shadow cone but it is then strongly modified by blocking along the outgoing trajectory.

Collecting experimental ICISS data across a large range of scattering angles to measure how the features change is time consuming, however, and too high of an ion beam fluence would damage the surface. Instead, simulations are performed here as a function of scattering angle to verify the nature of the flux peak features. This is a new approach to the analysis of large-angle ion scattering data that reveals much more information than can be gleaned solely from simulations of ICISS scans at fixed scattering angles.

Figure 5 shows a three-dimensional (3D) representation of calculated ICISS polar scans for scattering angles ranging from 100º to 180º. If Fig. 5 is cut along scattering angle 161º, for example, the data would be the same as the ICISS polar scan collected using a scattering angle of 161º, as shown in Fig. 2. Figure 5 can be used to clearly identify the trajectory responsible for each feature, and they are marked with the same labels used in Fig. 2.

Some of the features are due purely to either shadowing or blocking. Figure 5 clearly shows that the positions of features 1s and 4s are independent of scattering angle, which confirms that they result from shadowing, and are thus features that can be analyzed completely within the basic ICISS protocol. Feature 6b shifts linearly with scattering angle, confirming that it is purely due to blocking, which also enables a simple analysis.

Feature 2sb involves both shadowing and blocking. The position of feature 2sb does not shift with scattering angle, showing that shadowing is involved, but the intensity changes with scattering angle in a manner that indicates a contribution from blocking. For example, feature 2sb has zero intensity in the region around a 150º scattering angle which corresponds to ions that have made a hard collision with the second layer Bi atoms but are then completely blocked from reaching the detector by the first layer Se atoms above. If there were Se vacancies in the first layer, then the intensity of feature 2sb would not go to zero near a 150º scattering angle. In addition, a cut of the bottom panel of Fig. 5 along a 17° incident polar angle shows enhanced intensity at scattering angles 130º and 180º. These maxima in the Bi SSP intensity indicate the angular positions of the edges of the blocking cone. This analysis shows that feature 2sb does involve both shadowing and blocking and further confirms that the three-atom trajectory for 2sb drawn in Fig. 4 is correct. Such a conclusion is difficult to make directly from experimental data at a single scattering angle, but the use of simulations as a function of scattering angle makes it possible.

Note that the intensity of feature 1s also changes with scattering angle, but the change does not have to do with blocking. The maximum in 1s at the lowest scattering angles occurs because the impact parameter and the scattering cross sections generally increase at smaller angles. Maxima at 180° also occur for all single scattering trajectories due to a combined shadowing and blocking enhancement, as explained below. Also, there are no other maxima in the intensity of 1s as a function of scattering angle that would indicate the edge of a blocking cone.

The incident polar angle of feature 5s is fairly constant with scattering angle, indicating that the main contribution is from shadowing, but the intensity vs. scattering angle has a small local minimum, suggesting that some blocking is involved. The blocking is due to a first layer Se atom and occurs primarily between 100° and 140º, however, so that it is absent when the scattering angle reaches the 161º used in the experiments. Thus, it is marked here as 5s instead of 5sb.

The above discussion shows that blocking makes data analysis more complicated, so it can sometimes be useful for the scattering angle to be as close to 180º as possible. Figure 5 shows that when the scattering angle is exactly 180º, as in CAICISS [25], all of the features due to blocking disappear. In addition, the intensities of all features at a 180º scattering angle are enhanced because the same atom that forms a shadow cone to focus the incoming ion beam also forms a blocking cone that focuses the outgoing beam, and the edges of the shadow and blocking cones of the same atom pair overlap so that the scattered intensity is enhanced even further. The enhancement in the intensity of features 1s, 4s and 5s at scattering angles close to 180º is due to this effect instead of blocking. Note that the enhancement of 1s is the most pronounced as it is the SFP and results from atom pairs that are farther apart than the other atom pairs, which increases the intensity of this shadowing-blocking focusing effect. The enhancement for 2sb at 180º scattering angle is due to both this effect and blocking.

Although features due to both shadowing and blocking complicate the data analysis, once their origin is identified by the method presented here, they can provide better resolution in determining structural parameters because of their sensitivity to the atomic positions. For example, when the spacing between the first and second layers is increased in the simulations from 1.55 Å to 1.66 Å, feature 2sb completely disappears at a 161º scattering angle, which agrees with the result from Ref. [11] in which the SSP intensity was monitored as a function of azimuthal angle. This example demonstrates the usefulness of analyzing such complex features with the use of simulations that vary two independent angles in a structural determination performed via large scattering angle LEIS angular distributions.

**IV. Conclusions**

ICISS angular scan experiments and molecular dynamics simulations are performed for $Bi_2Se_3$ along the [120] and [$\bar{1}\bar{2}0$] azimuths. The match of all of the flux peak features between experimental and calculated ICISS data at a 161° scattering angle shows that the IBA-prepared surfaces are QL-terminated. Calculated polar scans performed over all scattering angles are further employed to accurately identify the trajectories responsible for each of the flux peaks in the experimental data. With the help of such simulations, the trajectories responsible for the features are much better understood, and the identifications shown in Fig. 3 are confirmed. This includes common two-atom flux peaks as well as a novel trajectory that involves interaction of the projectile with three target atoms. Also, comparisons of the experimental data to simulations have helped identify a minimum in the multiple scattering background that could be misinterpreted as a flux peak in an ICISS scan. This work outlines a new approach in which simulations over a large range of scattering angles are used to unambiguously identify the trajectories in large angle LEIS angular scans.

**V. Acknowledgements**

This material is based on work supported by, or in part by, the U.S. Army Research Laboratory and the U.S. Army Research Office under Grant No. 63852-PH-H.

# References

1. M. Aono and R. Souda, Jpn. J. Appl. Phys. **24**, 1249 (1985).

2. J. A. Yarmoff and R. S. Williams, Surf. Sci. **166**, 101 (1986).

3. Th. Fauster, Vacuum **38**, 129 (1988).

4. J. B. Pendry, *Low Energy Electron Diffraction: The Theory and Its Application to Determination of Surface Structure* (Academic Press, London, 1974).

5. C. S. Fadley, in *Synchrotron Radiation Research: Advances in Surface and Interface Science Techniques*, edited by R. Z. Bachrach (Springer US, Boston, MA, 1992), p. 421.

6. G. Binnig, H. Rohrer, C. Gerber, and E. Weibel, Phys. Rev. Lett. **49**, 57 (1982).

7. H. Niehus, W. Heiland, and E. Taglauer, Surf. Sci. Rep. **17**, 213 (1993).

8. W. J. Rabalais, *Principles and applications of ion scattering spectrometry : surface chemical and structural analysis* (Wiley, New York, 2003).

9. W. Zhou, H. Zhu, and J. A. Yarmoff, Phys. Rev. B **94**, 195408 (2016).

10. W. Zhou, H. Zhu, C. M. Valles, and J. A. Yarmoff, Surf. Sci., submitted.

11. W. Zhou, H. Zhu, and J. A. Yarmoff, J. Vac. Sci. Technol. B **34**, 04J108 (2016).

12. W. Zhou, H. Zhu, and J. A. Yarmoff, Phys. Rev. Lett., submitted.

13. X.-L. Qi and S.-C. Zhang, Rev. Mod. Phys. **83**, 1057 (2011).

14. Z. Wei, Y. Rui, Z. Hai-Jun, D. Xi, and F. Zhong, New J. Phys. **12**, 065013 (2010).

15. Y. Xia, D. Qian, D. Hsieh, L. Wray, A. Pal, H. Lin, A. Bansil, D. Grauer, Y. S. Hor, R. J. Cava, and M. Z. Hasan, Nat. Phys. **5**, 398 (2009).

16. R. W. G. Wyckoff, *Crystal Structures - Volume 2 : Inorganic Compounds $RX_n$, $R_nMX_2$, $R_nMX_3$* (Interscience Publishers, New York, 1964).

17. H. Lin, T. Das, Y. Okada, M. C. Boyer, W. D. Wise, M. Tomasik, B. Zhen, E. W. Hudson, W. Zhou, V. Madhavan, C.-Y. Ren, H. Ikuta, and A. Bansil, Nano Lett. **13**, 1915 (2013).

18. B. Yan, D. Zhang, and C. Felser, Phys. Status Solidi RRL **7**, 148 (2013).

19. X. Wang and T. C. Chiang, Phys. Rev. B **89**, 125109 (2014).

20. Y. N. Zhang, J. Chem. Phys. **143**, 151101 (2015).

21. M. A. Karolewski, Nucl. Instr. Meth. Phys. Res. B **230**, 402 (2005).

22. D. D. dos Reis, L. Barreto, M. Bianchi, G. A. S. Ribeiro, E. A. Soares, W. S. Silva, V. E. de Carvalho, J. Rawle, M. Hoesch, C. Nicklin, W. P. Fernandes, J. Mi, B. B. Iversen, and P. Hofmann, Phys. Rev. B **88**, 041404 (2013).

23. G. E. Shoemake, J. A. Rayne, and R. W. Ure, Phys. Rev. **185**, 1046 (1969).

24. O. S. Oen, Surf. Sci. Lett. **131**, L407 (1983).

25. M. Aono, M. Katayama, and E. Nomura, Nucl. Instr. Meth. Phys. Res. B **64**, 29 (1992).

26. D. S. Gemmell, Rev. Mod. Phys. **46**, 129 (1974).

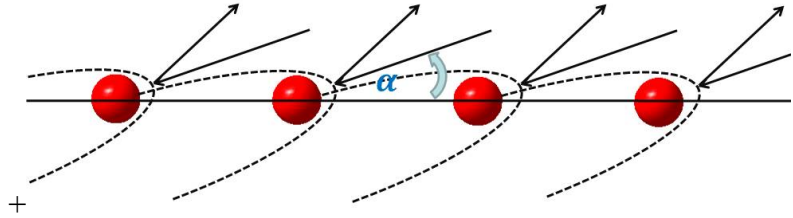

**Figure 1.** A schematic illustrating the geometry for flux peak formation in ICISS. The arrows illustrate projectile trajectories at a 161° scattering angle, α is the angle of the incoming ions with respect to the atomic chain, and shadow cones are illustrated by dashed lines. Atoms in a chain interact with the edges of the shadow cones formed by adjacent atoms.

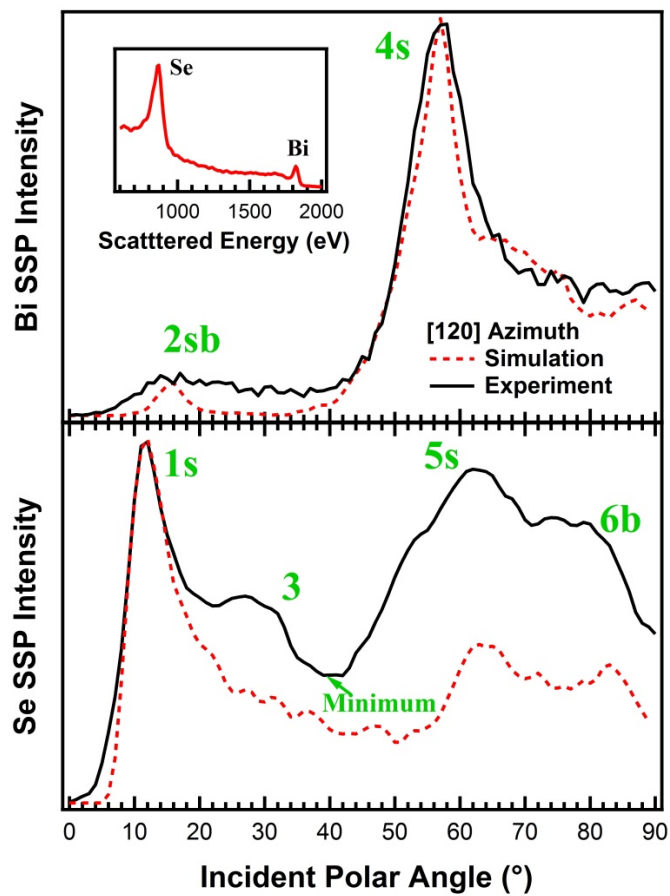

**Figure 2.** Experimental and calculated ICISS polar scans of the Bi and Se SSPs for 3 keV Na$^+$ scattering from an IBA-prepared Bi$_2$Se$_3$ surface collected along the [120] azimuth using a scattering angle of 161º. The incident polar angle is given with respect to the surface plane. The experimental and simulated data are adjusted to have the same maximum intensity. Each prominent feature is labeled (see text). The inset in the upper panel is an energy spectrum collected using an incident polar angle of 12º along the [120] azimuth with a 161º scattering angle.

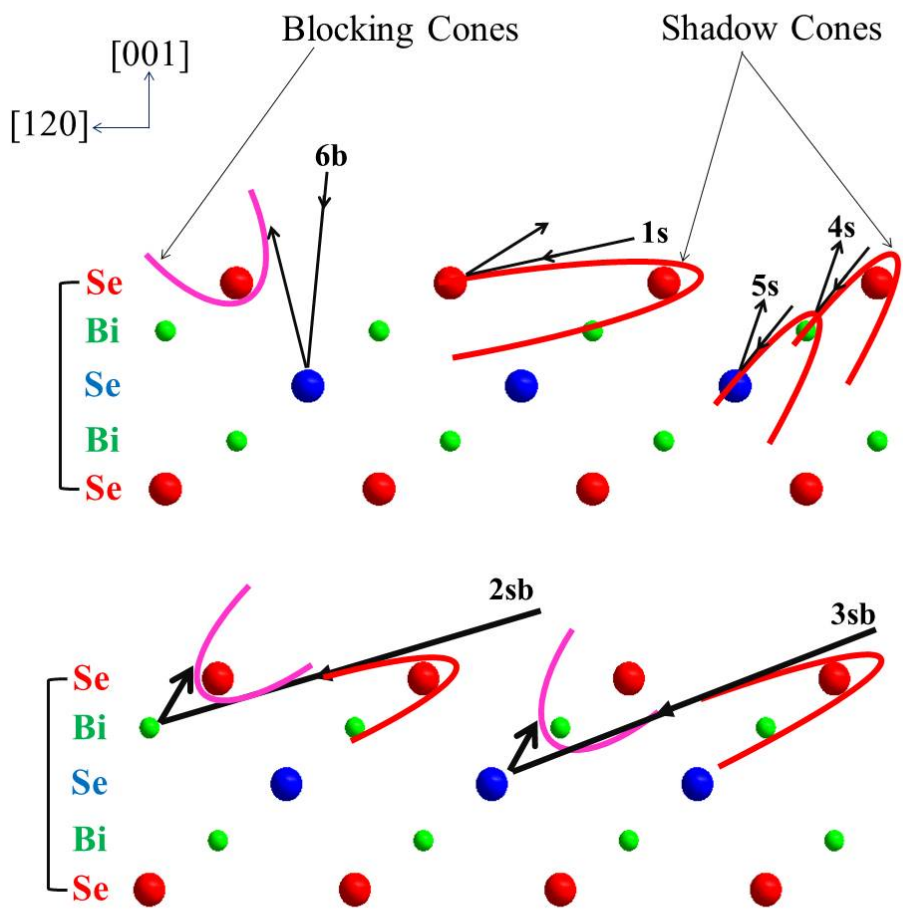

**Figure 3.** Side view schematic diagram of the ($\bar{1}$20) plane showing the atoms pairs that contribute to the features in Fig. 2. The trajectories are labeled with a number and 's' or 'b' characters representing whether shadow or blocking cones are involved in creating the associated flux peak. The sizes of the atoms are represented by their ionic radii, while the shadow and blocking cones are not drawn to scale.

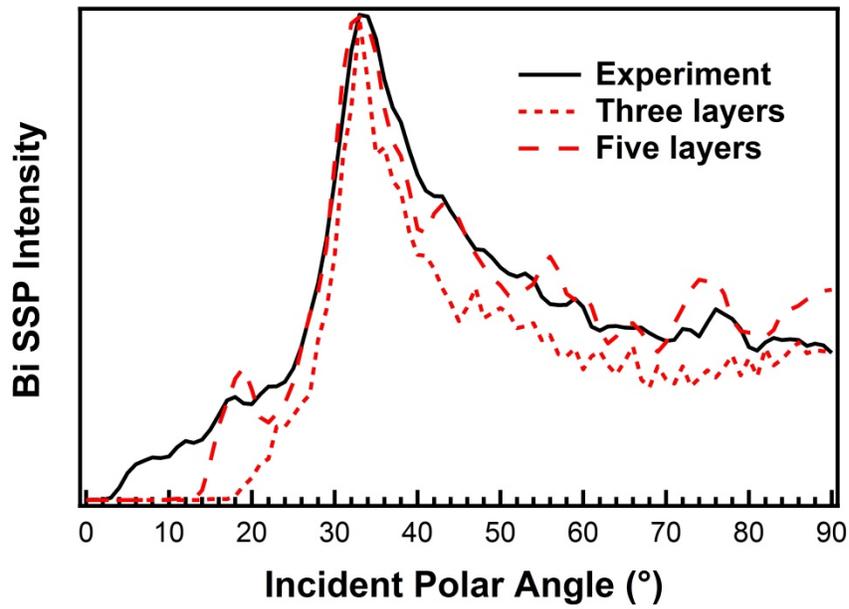

**Figure 4.** ICISS polar scan of the Bi SSP for 3 keV Na[+] scattering from an IBA-prepared $Bi_2Se_3$ surface collected along the $[\bar{1}\bar{2}0]$ azimuth shown along with the results of simulations. Two different bulk-terminated targets are used in the simulations. One has three atomic layers (short dashed line), while the other has five layers (long dashed line).

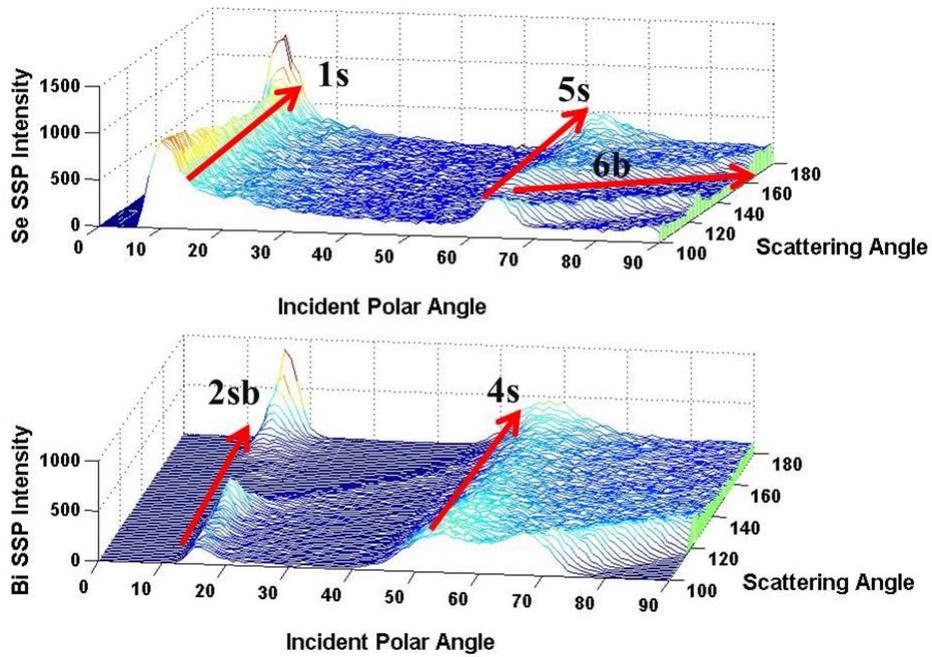

**Figure 5.** Three-dimensional representation of calculated Se SSP and Bi SSP intensities for 3 keV Na$^+$ scattering from bulk-terminated Bi$_2$Se$_3$ as functions of the incident polar angle along the [120] azimuth and the scattering angle ranging from 100° to 180º.